\documentclass[11pt]{article}
\usepackage[utf8]{inputenc}
\usepackage{graphicx}
\usepackage{amsmath, amssymb}
\usepackage{authblk}  % for author + affiliation
\usepackage{cite}     % compact numbered citations
\usepackage{geometry} % better page margins
\usepackage{caption}
\usepackage{hyperref} % for clickable links and DOIs

\title{Machine Learning-Driven Compensation for Non-Ideal Channels in AWG-Based FBG Interrogator}

\author[1,2]{Ivan A. Kazakov}
\author[4,2]{Iana V. Kulichenko}
\author[1,2]{Egor E. Kovalev}
\author[1,2]{Angelina A. Treskova}
\author[3,2]{Daria D. Barma}
\author[1,2]{Kirill M. Malakhov}
\author[3]{Ivan V. Oseledets}
\author[1,2]{Arkady V. Shipulin}

\affil[1]{Photonic Integration Research Lab, Skolkovo Institute of Science and Technology, Moscow, Russia}
\affil[2]{FiberPipe LLC, Moscow, Russia}
\affil[3]{Material Science Lab, Skolkovo Institute of Science and Technology, Moscow, Russia}
\affil[4]{Artificial Intelligence Center, Skolkovo Institute of Science and Technology, Moscow, Russia}

\date{June 15, 2025}

\begin{document}

\maketitle

\begin{abstract}
We present an experimental study of a fiber Bragg grating (FBG) interrogator based on a silicon oxynitride (SiON) photonic integrated arrayed waveguide grating (AWG). While AWG-based interrogators are compact and scalable, their practical performance is limited by non-ideal spectral responses. To address this, two calibration strategies within a 2.4 nm spectral region were compared: (1) a segmented analytical model based on a sigmoid fitting function, and (2) a machine learning (ML)-based regression model. The analytical method achieves a root mean square error (RMSE) of 7.11 pm within the calibrated range, while the ML approach based on exponential regression achieves 3.17 pm. Moreover, the ML model demonstrates generalization across an extended 2.9 nm wavelength span, maintaining sub-5 pm accuracy without re-fitting. Residual and error distribution analyses further illustrate the trade-offs between the two approaches. ML-based calibration provides a robust, data-driven alternative to analytical methods, delivering enhanced accuracy for non-ideal channel responses, reduced manual calibration effort, and improved scalability across diverse FBG sensor configurations.
\end{abstract}

\noindent\textbf{Keywords:} Fiber Bragg Grating, Arrayed Waveguide Grating, Machine Learning, FBG interrogator, interrogation function, algorithm for interrogator

% Optional note
\noindent\textbf{Note:} This is a preprint submitted to arXiv. The manuscript has also been submitted to IEEE Sensors Letters and is currently under peer review.

\section{Introduction}
\label{sec:introduction}

Fiber Bragg gratings (FBGs) enable precise sensing of deformation, temperature, and vibrations in critical infrastructure \cite{ref_1,ref_2}. These sensors are used in pairs with interrogators capable of detecting sub-nanometer shifts in the Bragg wavelength. Two dominant architectures exist: (i) tunable laser-based systems offer high precision and optical power \cite{ref_17}, but there is a speed process limitation, and (ii) spectrometer-based systems provide high acquisition rates, but they are constrained by optical power and system size \cite{ref_3,ref_4}.

Photonic integrated circuits (PIC) with arrayed waveguide gratings (AWG) present a scalable alternative for replacing bulky spectrometers and enabling MHz-class sampling with fast photodiodes \cite{ref_15, ref_14, ref_5}. However, real AWG channels deviate from ideal Gaussian shapes, causing asymmetry and ripple. These deviations, along with practical factors such as SLED output fluctuations due to thermal or environmental noise \cite{ref_2}, reduce the accuracy of common centroid or peak-ratio methods \cite{ref_7}. Recent work has focused on optimizing AWG layout and fabrication to reduce such effects \cite{ref_8, ref_9}, but imperfections remain a challenge for robust demodulation.

To compensate for such imperfections, previous work employed polynomial fits \cite{ref_10} or auxiliary interferometric paths \cite{ref_11}, achieving picometer-scale resolution at the cost of complexity or footprint.

\begin{figure}[h]
\centerline{\includegraphics[width=0.5\columnwidth]{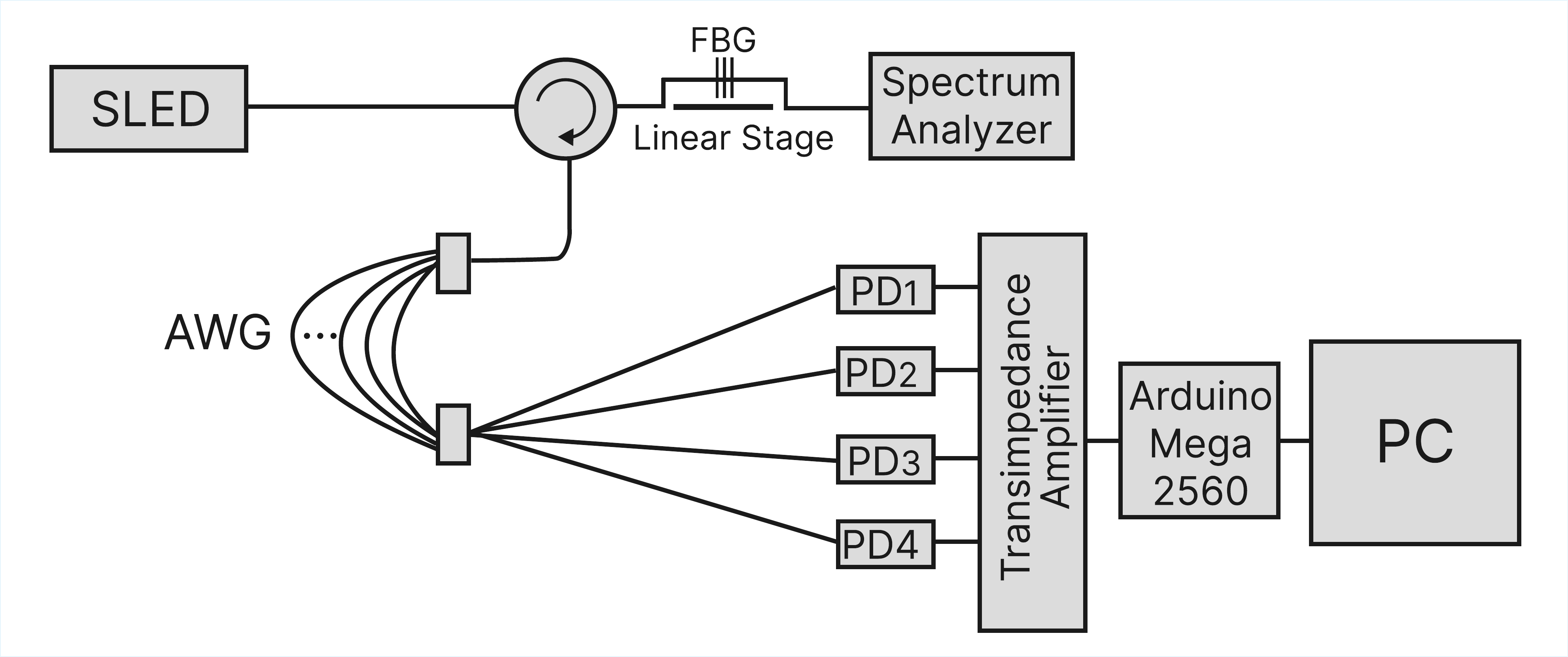}}
\caption{Scheme of developed interrogator. SLED - broadband superluminescent diode, PD - photodiodes, PC - computer.}
\label{fig:OI_scheme}
\end{figure}

In this work, we experimentally evaluate an AWG-based FBG interrogator and compare two calibration approaches within the 2.4 nm spectral region between four AWG channels: (i) an analytical model based on a segmented sigmoid fitting applied to each channel transition zone, and (ii) a machine learning (ML) regression model trained across all available channel data.

The analytical method achieves a root mean square error (RMSE) of 7.11 pm, while the ML model reduces it to 3.17 pm and additionally generalizes over an extended 2.9 nm wavelength span with sub-5 pm accuracy, eliminating the need for local re-fitting.

ML-based calibration provides a data-driven alternative to analytical methods, enabling improved accuracy under non-ideal channel responses, reduced dependence on manual fitting procedures, and enhanced scalability across various FBG sensor types and deployment conditions.

\section{Methods}

\label{sec:methods}
The section describes the experimental setup, including the interrogator prototype and the algorithms used to collect the data for calibration and measurement.

\subsection{Hardware}
\label{ssec:hardware}

During the research, an experimental interrogator setup was assembled and calibrated. Figure \ref{fig:OI_scheme} shows the interrogator scheme, consisting of: a broadband superluminescent diode (SLED, RZSLD-1550-20-BS-FA-14-HC, 20 mW), an optical circulator, a FBG ($\lambda\approx$ 1539 nm, FWHM = 0.2 nm), a 16-channel SiON-based AWG demultiplexer (FOTIS LLC, temperature stabilized with TEC\cite{ref_6}), photodiodes with transimpedance amplifiers, and an Arduino Mega2560 for data acquisition.

During the experiment, the FBG was mounted on a stage for wavelength tuning via strain. Reflected light was routed through the AWG, and channel outputs were averaged over 100 samples. An Anritsu MS9740B optical spectrum analyzer was used for reference FBG wavelength measurements.

\begin{figure}[h]
\centerline{\includegraphics[width=0.5\columnwidth]{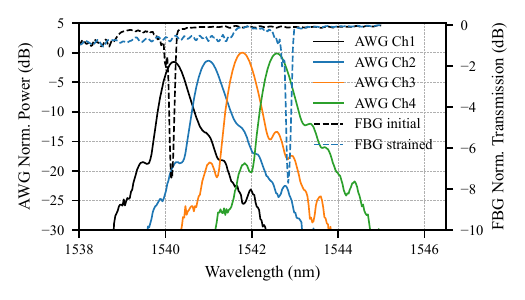}}
\caption{Experimental transmission spectra of four adjacent AWG channels (AWG Ch1 - AWG Ch4), the initial and final strain-tuned reflection spectra of the operated FBG in the study.}
\label{fig_awg}
\end{figure}

Data acquisition was performed over multiple measurement sessions by manually tuning the FBG back and forth across the AWG spectral range. Therefore, wavelengths in the dataset were not uniformly spaced.  This process yielded 438 samples in the range of 2.9 nm.

\begin{figure}[h]
\centerline{\includegraphics[width=0.5\columnwidth]{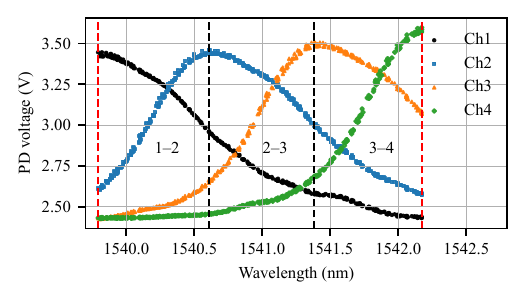}}
\caption{Photodiode voltage response across the scanned wavelength range for four AWG channels. The red vertical lines indicate the boundaries of the 2.4 nm operating range within which both the analytical and ML-based algorithms were evaluated. The black vertical lines denote the segmentation thresholds used by the analytical method to identify the active region for each channel pair (1–2, 2–3, or 3–4), based on precomputed photodiode ratio conditions.}
\label{fig_cal_data}
\end{figure}

To assess generalization, we used both random (80\%/20\%, 341 samples, fixed seed) and chronological (259 train samples, 82 test samples) train/test splits. The dataset was limited to a 2.4 nm range. Figure~\ref{fig_cal_data} illustrates the photodiode voltage response as a function of wavelength. The signal shapes define the operating region and transition boundaries, which are used in our experiments.

\subsection{Analytical Algorithm}
\label{sec:anal_methods}

The AWG-based interrogation algorithm is based on segmented calibration, extending the method originally introduced in \cite{ref_12} for two AWG channels. In this work, we generalize it to four channels and introduce a range detection mechanism to achieve a transition between regions.

The method operates in two stages: (1) detection of the active spectral region (i.e., the pair of adjacent AWG channels where the FBG reflection occurs), and (2) interpolation of the Bragg wavelength using a pre-fitted sigmoid function for that region.

Two threshold ratios are precomputed from the calibration dataset to detect the active region. These thresholds are based on voltage ratios between specific photodiode channels:
\begin{itemize}
    \item \(P_3 / P_1\), used to separate regions 1–2 and 2–3;
    \item \(P_4 / P_2\), used to separate regions 2–3 and 3–4.
\end{itemize}

Based on voltages \((P_1 - P_4)\), the algorithm selects one of three overlapping channel pairs: 1–2, 2–3, or 3–4. Once the region is detected, the corresponding calibration model is applied.

The wavelength demodulation relies on the voltage ratio $(P_{i+1}/P_{i})$ between adjacent channels, which is mapped to wavelength using a four-parameter sigmoid function (Eq. \ref{eq:fitting_function}):

\begin{equation}
   f(x) = \frac{-L}{1 + \exp(-k(x - x_0))} + b
\label{eq:fitting_function}
\end{equation}

Here, \(x\) is the Bragg wavelength and \(f(x)\) is the measured voltage ratio. The parameters \((L, k, x_0, b)\) are fitted individually for each channel pair.

\subsection{ML Methods}
\label{sec:ml_methods}

FBG wavelength demodulation employing an AWG can be framed as a supervised regression problem, aiming to estimate the Bragg reflection wavelength $\lambda$ using the optical power vector $\mathbf{P} = [P_1, P_2, P_3, P_4]^\top$.

\subsubsection{Regression Models}

To learn the mapping from the detected optical powers to the Bragg wavelength, a set of regression models was implemented and optimized: Exponential Ridge Regression, Random Forest, CatBoost,  Gradient Boosting Regression, Polynomial Ridge Regression, Support Vector Regression, and XGBoost. Each model was processed through a scikit-learn pipeline: a custom transformer first generated polynomial, logarithmic, exponential, and square-root features to capture non-linear behavior, after which StandardScaler normalized every feature to zero mean and unit variance, ensuring balanced scales and stable optimization.

\subsubsection{Training Process}
The feature matrix $\mathbf{X} \in \mathbb{R}^{N \times 4}$ is derived from the normalized optical power values $[P_1, P_2, P_3, P_4]$. The target vector $\mathbf{y} \in \mathbb{R}^{N}$ is comprised of the corresponding Bragg reflection wavelengths $\lambda$. Final model selection was based on the lowest test set RMSE among all trained candidates.

Hyperparameter tuning was performed using the Optuna framework \cite{ref_13}, which employs TPE-based Bayesian optimization with early stopping to efficiently explore complex parameter spaces, achieving faster convergence and higher accuracy than traditional grid or random search. Each model was optimized to minimize 5-fold cross-validated MSE using repeated 3-fold CV (2 repeats) with a fixed seed for robustness. The search space was limited to 30 trials to balance accuracy and computational cost.

\section{Results}
A comparative evaluation of the same experimental dataset was performed. The calibration dataset was acquired over several hours by gradually stretching the FBG while recording voltages from photodiodes. Then, two interrogation approaches were applied to this dataset: the analytical sigmoid-based calibration method and the ML-based regression algorithms. 

As Table~\ref{tab:ml_model_results} shows, exponential ridge regression gives the lowest error on the random 80/20 split (RMSE = 3.17 pm), slightly ahead of the analytical method (7.11 pm). In the time-ordered split (Table~\ref{tab:model_results}), the analytical model is better (RMSE = 5.39 pm vs.\ 12.72 pm). All \(R^{2}\) scores hover near 1, which is expected: over the narrow 2.4 nm range, picometre-scale errors contribute only a minute fraction of the total variance.

Exponential regression was selected as the best-performing ML model due to its high accuracy (lowest RMSE and MAE). Its simple functional form is well-suited for real-time deployment, especially in resource-constrained settings. Also, ridge works well because the expanded features make the power–wavelength map nearly linear, while its penalty suppresses collinearity and overfitting.

In summary, ML models outperformed the analytical method when the test data covered the full operational range, making them well-suited for controlled environments. The analytical approach generalizes better to unseen data and is more robust to distribution shifts. ML models, however, scale more easily to additional channels and complex response patterns.

\begin{table}[t]
\centering
\caption{Comparison of Regression Models (Random splitting)}
\label{tab:ml_model_results}
\renewcommand{\arraystretch}{1.6}
\resizebox{0.5\textwidth}{!}{
\begin{tabular}{|p{11em}|c|c|c|c|}
\hline
\textbf{Model} & \textbf{MAE(pm)} & \textbf{RMSE(pm)} & \textbf{\boldmath$R^2$} \\
\hline
\multicolumn{1}{|p{11em}|}{\textbf{Exponential Ridge Regression}}         & \textbf{2.61} & \textbf{3.17} & \textbf{0.99998} \\
\multicolumn{1}{|p{8em}|}{Polynomial Ridge Regression} & 2.83 & 3.52 & 0.99998 \\
\multicolumn{1}{|p{11em}|}{\textbf{Sigmoid (Analytical)}} & \textbf{5.76} & \textbf{7.11} & \textbf{0.99990} \\
\multicolumn{1}{|p{11em}|}{Support Vector Regression}     & 6.49 & 9.12 & 0.99984 \\
\multicolumn{1}{|p{11em}|}{Random Forest}                 & 7.94 & 9.70 & 0.99982 \\
\multicolumn{1}{|p{11em}|}{Gradient Boosting}             & 8.06 & 10.21 & 0.99980 \\
\multicolumn{1}{|p{11em}|}{XGBoost}                       & 10.69 & 14.22 & 0.99961 \\
\multicolumn{1}{|p{11em}|}{CatBoost}                      & 12.37 & 14.89 & 0.99958 \\
\hline
\end{tabular}
}
\end{table}

\begin{table}[t]
\centering
\caption{Comparison Of ML And Analytical Methods}
\label{tab:model_results}
\renewcommand{\arraystretch}{1.6}
\resizebox{0.5\textwidth}{!}{
    \begin{tabular}{|c|c|c|c|}
    \hline
    \textbf{Type of dataset splitting} & \textbf{MAE (pm)} & \textbf{RMSE (pm)} & $\mathbf{R^2}$ \\
    \hline
    \multicolumn{4}{|c|}{\textbf{ML (Exponential)}} \\
    \hline
    Random (80\%~/~20\%) & 2.612 & 3.171 & 0.999981 \\[0.2cm]
    \hline
    Sequential & 10.634 & 12.719 & 0.999697 \\[0.2cm]
    \hline
    \multicolumn{4}{|c|}{\textbf{Analytical}} \\
    \hline
    Random (80\%~/~20\%) & 5.758 & 7.106 & 0.999903 \\[0.2cm]
    \hline
    Sequential & 4.646 & 5.392 & 0.999945 \\[0.2cm]
    \hline
    \end{tabular}
}
\end{table}

\begin{figure}[h]
\centering
\includegraphics[width=0.5\columnwidth]{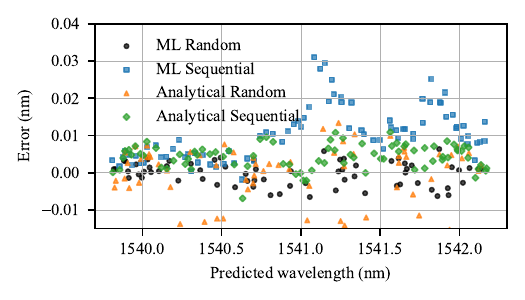}
\caption{Errors of the best ML model (Exponential Regression) and analytical model as functions of predicted wavelength.}
\label{fig:errorplot}
\end{figure}

To assess prediction behavior, we analyzed the residuals of the best machine learning (ML) model (Exponential Regression) and the analytical model as functions of the predicted wavelength (Fig. \ref{fig:errorplot}). The errors were quantified using the signed error $error = \lambda_{pred} - \lambda_{ref}$ metric. Errors are centered around zero with no visible trend, indicating good generalization. Most fall below 15 pm, with a few outliers approaching 30 pm.

\section{Discussion}
\label{sec:discussion}

Table~\ref{tab:comparison} compares the proposed approach with recent FBG demodulation methods. The analytical sigmoid model achieved a mean absolute error (MAE) of 5.76 pm, while the ML-based exponential regression improved it to 2.61 pm.

Compared to polynomial-calibrated AWGs \cite{ref_14,ref_15}, the ML method simplifies deployment by avoiding per-channel curve fitting. It also matches the accuracy of resonant MRR-based interrogators \cite{ref_16} while offering greater scalability. Unlike high-speed FDML laser systems \cite{ref_17}, the AWG+ML approach is cost-effective and enables better integration on PICs.

However, the ML performance relies on the quality of calibration data. Additionally, ML models may degrade under environmental drift or unseen sensor aging effects. Domain adaptation or continual learning strategies could address such limitations in future implementations.

The analytical method, based on voltage ratios, is inherently robust to potential effects of SLED output fluctuations, such as minor intensity drift. In our measurements, SLED instability remained within 0.1 dB, negligible compared to the 20 dB system dynamic range. In contrast, ML models may be more sensitive to such variations, which could contribute to reduced performance under time-separated testing.

In future work, it is planned to: (i) embed the ML model in microcontroller firmware for real-time demodulation; (ii) scale the demonstrator to 16 and 44 channels for multi-FBG sensing; and (iii) apply domain adaptation to improve robustness under varying environmental conditions.

\begin{table}[h]

\caption{Comparison with Recent FBG Interrogators (2020–2025)}
\renewcommand{\arraystretch}{1.6}
\label{tab:comparison}
\centering
\resizebox{0.5\textwidth}{!}{
\begin{tabular}{|c|c|c|c|}
\hline
\multicolumn{1}{|p{10em}|}{\textbf{Method}} & \multicolumn{1}{|p{3em}|}{\textbf{MAE (pm)}} &\multicolumn{1}{|p{3em}|}{\textbf{ Speed (Hz)}}& \multicolumn{1}{|p{5em}|}{\textbf{Dyn.~Range (dB)}} \\
\hline
\multicolumn{1}{|p{10em}|}{AWG + analytical} & 5.76  & 100   & 20 \\
\multicolumn{1}{|p{10em}|}{AWG + ML }  & 2.61  & 100   & 20\\
\multicolumn{1}{|p{10em}|}{Silica PLC AWG\cite{ref_14} } & $<1$  &    –   &  – \\
\multicolumn{1}{|p{10em}|}{Si AWG CoG \cite{ref_15}} & 0.73 &    –   &  25 \\
\multicolumn{1}{|p{10em}|}{MRR \cite{ref_16}} & 10–30 & 100\,k & 10\\
\multicolumn{1}{|p{10em}|}{FDML swept‐laser\cite{ref_17}} & 1-10  & 10-100\,k & 40 \\
\hline
\end{tabular}
}
\end{table}

\section{Conclusion}
In this work, two calibration algorithms for a photonic AWG-based FBG interrogator were developed and benchmarked. The analytical 4-parameter sigmoid model achieved $\approx$7~pm RMSE, while the ML-based polynomial regression reached $\approx$3~pm. 

ML methods outperform analytical calibration due to their adaptability to diverse hardware setups and operating conditions. While analytical models become increasingly complex with more sensors and shifting environments, ML approaches scale efficiently by learning directly from data, making them robust and effective for practical deployment.

\section*{Acknowledgment}
The authors express their gratitude to the academic support of Skolkovo Institute of Science and Technology (Skoltech), Zelenograd Nanotechnological Center (ZNTC), and LLC "FOTIS" for their collaboration. The authors would like to express their gratitude to Andrey Gorelov from Photonic Integration Research Lab, Skoltech, for his assistance in experimenting, to Petr Krotov for support with the data acquisition, and to Nika Oktyabr for the preparation of the graphical abstract.

\end{document}